\documentclass[twocolumn,showpacs,floatfix,aps,prl]{revtex4}

\usepackage[dvips]{graphicx,color}
\usepackage{dcolumn}
\usepackage{bm}
\usepackage{amsmath, amssymb, mathrsfs}

\newcommand{\rrh}{\hat{\mathbf{r}}}
\newcommand{\rr}{\mathbf{r}}
\newcommand{\kk}{\mathbf{k}}
\newcommand{\KK}{\mathbf{K}}

\newcommand{\qq}{\mathbf{q}}
\newcommand{\RR}{\mathbf{R}}
\newcommand{\be}{\begin{equation}}
\newcommand{\ee}{\end{equation}}
\newcommand{\bea}{\begin{eqnarray}}
\newcommand{\eea}{\end{eqnarray}}
\def\epsb{\mbox{\boldmath$\epsilon$}}
\def\epsbs{\mbox{\boldmath\scriptsize$\epsilon$}}
\def\cheta{\check}

\begin{document}

\title{Spectrum of Light in a Quantum Fluctuating Periodic Structure}

\author{Mauro Antezza and Yvan Castin}
\affiliation{Laboratoire Kastler Brossel, \'{E}cole Normale Sup\'{e}rieure, CNRS and UPMC, Paris, France}

\date{\today}

\begin{abstract}
We address the general problem of the excitation spectrum for light coupled to 
scatterers having quantum fluctuating positions around the sites of a periodic
lattice.
In addition to providing an imaginary part to the spectrum, we show that
these quantum fluctuations affect the real part of the spectrum,
in a way that we determine analytically.
Our predictions may be observed with ultracold atoms in an optical lattice,
on a $J=0\to J'=1$ narrow atomic transition.
As a side result, we resolve a controversy for the occurrence of a spectral gap
in a fcc lattice.
\end{abstract}


\maketitle

The investigation of light propagation in a periodic structure
is a fundamental problem in condensed matter physics \cite{Hopfield58,Agranovich60}, 
ranging from the physics of photonic crystals \cite{FCBook} to X ray \cite{Pastori} 
or even $\gamma$ ray \cite{Kagan,gamma_exp}
scattering by a crystal. It recently gained a renewed interest 
\cite{Lagendijk96,Lagendijk98,Knoester06,Carusotto08,Cirac08,Zoubi09}
thanks to the
possibility of producing artificial periodic structures of 
quantum dots \cite{Knoester06} or of atoms \cite{Bloch}, 
which have a much larger spatial period
than natural crystals, allowing a measurement with a laser of the excitation spectrum 
over the whole Brillouin zone.

In reality, strictly periodic structures do not exist: the quantum (if not thermal)
fluctuations of the positions of the scatterers in the structure are unavoidable.
From general grounds,  we know that these fluctuations affect the spectrum of light 
in two ways. First, they introduce a {\sl dissipative} component: the elementary excitation
spectrum acquires an imaginary part, a well known effect of phonon coupling
in condensed matter physics.
Second, they introduce a {\sl reactive} component,
modifying the real part of the excitation spectrum. Whereas the fluctuations
of the scatterers positions have indeed been taken into account
in simplified models \cite{Zoubi09}, the only explicit prediction
for the corresponding modification to the excitation spectrum
was given in \cite{Kagan}.

The problem however is not closed yet: in the limit of vanishing
position fluctuations, Eq.(4.9) of \cite{Kagan} for the real part of the
spectrum does {\sl not} reduce to the prediction given in \cite{Knoester06}
for {\sl fixed} scatterers, a prediction that also does {\sl not} coincide
with the one of \cite{Lagendijk96}.
This problem is not only formal, it may soon be addressed in current experiments
with narrow line cold atoms trapped in optical lattices 
\cite{Katori2008}, where measurements of the spectrum 
with a good precision may be performed.
Here we provide a conclusive analytical answer to this problem.


\noindent{\sl Model:}
Although our method to come is general,
we assume for concreteness that the scatterers are 
atoms coupled to the electromagnetic field
on an electronic transition between the ground state $g$ of 
spin $0$ and an excited state $e$ of spin $1$.
The atoms are trapped at the nodes of an optical lattice,
with nowhere more than one atom per site \cite{Bloch}.
In the deep-lattice limit,  tunneling is negligible and
the $i^{\rm th}$ atom is assumed   
to be harmonically trapped around lattice site $\RR_i$,
with the potential
$U_i(\rrh_i) = m\omega_{\rm ho}^2 (\rrh_i-\RR_i)^2/2.$
Here $m$ is the atomic mass, $\rrh_i$ is the position operator
of atom $i$, and $\omega_{\rm ho}$ is the atomic oscillation frequency.
In this regime, the fluctuations
of the atomic positions are purely on-site and uncorrelated, contrarily
to the case of phonons in a crystal \cite{Kagan}.
The Hamiltonian of our system \cite{Morice95} may be split in the non-interacting term $H_0$
and the atom-field dipolar coupling $V$,
$H = H_0 + V,$
with 
\bea
H_0 &=& \sum_{i=1}^{N} \left[\frac{\hat{\mathbf{p}}_i^2}{2m}
+U_i(\rrh_i)+ \sum_{\alpha} \hbar \omega_0 |i:e_\alpha\rangle \langle i:e_\alpha|\right]
\nonumber \\
&+&\int_{\mathcal{D}} 
d^3 k \sum_{\epsbs\perp\kk} \hbar c k\, \hat{a}^\dagger_{\kk\epsbs} \hat{a}_{\kk\epsbs}.
\label{eq:H0}
\eea
Here $N$ is the number of atoms, $\omega_0$ is the bare atomic resonance frequency,
the sum $\sum_\alpha$ over the three directions of space $x$, $y$, $z$ 
accounts
for the three-fold degeneracy of $e$,
$\mathcal{D}$ is the three-dimensional Fourier space truncated by a cut-off $k<k_M$,
and the annihilation and creation operators obey usual bosonic commutation relations
such as $[\hat{a}_{\kk\epsbs},\hat{a}^\dagger_{\kk'\epsbs'}]=
\delta_{\epsbs,\epsbs'} \delta(\kk-\kk')$, where $\epsb$ 
and $\kk$ are the photon polarization and wavevector.
The coupling $V$ is \cite{pas_delta}
\be
V = -\sum_{i=1}^{N} \hat{\mathbf{D}}_i \cdot  \hat{{\bf E}}_{\perp}(\rrh_i)
\ee
where  $\hat{D}_{i,\alpha} = d|i:e_\alpha\rangle\langle i:g| + \mbox{h.c.}$
is the component along direction $\alpha$ of the dipole operator $\hat{\mathbf{D}}_i$
of the $i^{\textrm{th}}$ atom,
proportional to the atomic dipole moment $d$, $\hat{{\bf E}}_{\perp}(\rr) =
\int_{\mathcal{D}} d^3k \sum_{\epsbs\perp\kk} \left[\mathcal{E}_k
\epsb\; \hat{a}_{\kk\epsbs}\; e^{i\kk\cdot \rr} + \mbox{h.c.}\right]$~is~the 
transverse electric field operator, 
and $\mathcal{E}_k=i (2\pi)^{-3/2}[\hbar kc/(2\varepsilon_0)]^{1/2}$.

\noindent{\sl Method:}
We treat the coupling $V$ to second order of perturbation theory,
to calculate atom-field elementary excitations mainly of atomic nature.
In practice for a lattice spacing
$\sim 1/k_0$, where $k_0=\omega_0/c$ is the resonant wavevector,
this requires $\Gamma\ll \omega_{\rm ho}$,
where  $\Gamma=d^2k_0^3/(3\pi\varepsilon_0\hbar)$ is  the free space atomic
spontaneous emission rate.
The energies of the system up to second order in $V$ are eigenvalues
of the effective Hamiltonian
\be
H_{\rm eff} = P H P + P V Q \frac{Q}{E^{(0)}Q-QH_0Q} QVP,
\label{eq:gen}
\ee
where $E^{(0)}$ is an eigenenergy of $H_0$, $P$ projects orthogonally 
onto the corresponding eigenspace of $H_0$, and $Q=I-P$.
We now obtain the excitation energies of the system
as the difference between excited states energies and the ground state energy.

For the perturbative calculation of the ground state energy $E_g$, we take
for $P$ the projector over the state with all the atoms in the electronic and motional
ground states, and the field in vacuum, so that $E_g^{(0)}=N \frac{3}{2}\hbar\omega_{\rm ho}$. 
We obtain $E_g=N \epsilon_g$, with \cite{neglige}
\be
\epsilon_g = \frac{3}{2}\hbar\omega_{\rm ho} 
-\frac{d^2}{\varepsilon_0}
\int_{\mathcal{D}} \frac{d^3k}{(2\pi)^3} \frac{\hbar c k}
{\hbar\omega_0+\hbar c k} \ .
\ee
For the excited state energies, we take for $P$ the projector $P_e$ over all the states
$||i:e_\alpha\rangle$,  where $||i:e_\alpha\rangle$ represents the
atom $i$ in the electronic state $e_\alpha$, the $N-1$ other atoms in the ground
electronic state, all the atoms in the motional ground state, 
and the field in vacuum. Then $P$ projects over a subspace of dimension $3N$.
Since this excited subspace is coupled by $V$ to the continuous part of the spectrum
of $H_0$, one has to replace $E^{(0)}$ by
$E_e^{(0)}+i 0^+$ in (\ref{eq:gen}), with $E_e^{(0)}=E_g^{(0)}+\hbar\omega_0$,
which gives rise to a complex excited state energy $E_e$, eigenvalue of
\begin{multline}
H_{\rm eff}^{e}  = [(N-1) \epsilon_g  + \epsilon_e ]P_e \\
+
\sum_{i\neq j} \sum_{\alpha,\beta}
\bar{g}_{\alpha\beta}(\RR_i-\RR_j) ||i:e_\alpha\rangle \langle j:e_\beta||.
\label{eq:heff}
\end{multline}
Here $\epsilon_e$ is the complex energy of a single atom \cite{neglige}
\be
\epsilon_e =  \hbar\omega_0 + \frac{3}{2}\hbar\omega_{\rm ho} + \frac{d^2}{3\varepsilon_0}
\int_{\mathcal{D}} \frac{d^3k}{(2\pi)^3} \frac{\hbar c k}{\hbar\omega_0+i0^+-\hbar c k}\ .
\label{eq:epse}
\ee
To obtain the excitation energies of the system we subtract the ground state energy
$E_g$ from (\ref{eq:heff}). In this subtraction, the dangerous terms proportional
to the number of atoms $N$ disappear and the interatomic distance independent
term gives the excitation energy for a single atom, that we split in a real part
and an imaginary part:
\be
\epsilon_e - \epsilon_g \equiv \hbar \omega_A - i\frac{\hbar \Gamma}{2}.
\ee
As expected, $\Gamma$ is the free space spontaneous emission rate,
and the effective atomic resonance frequency $\omega_A$ deviates
from $\omega_0$ by Lamb shift type terms.

Most interesting are the position dependent terms in (\ref{eq:heff}),
which contain
the effective coupling amplitude $\bar{g}_{\alpha\beta}(\rr)$
for the transfer of the atomic excitation
in between two different sites separated by $\rr$.
This coupling amplitude appears as the inverse Fourier transform
$\bar{g}_{\alpha\beta}(\rr)=
\int_{\mathcal D} [d^3k/(2\pi)^3] e^{i\kk\cdot\rr} \cheta{\bar{g}}_{\alpha\beta}(\kk)$
of the function
\be
\cheta{\bar{g}}_{\alpha\beta}(\kk) =  \frac{3\pi\hbar\Gamma}{k_0^3}  \ 
\frac{k^2\delta_{\alpha\beta}-k_\alpha k_\beta}{k_0^2-k^2+i0^+}
\ 
e^{-k^2 a_{\rm ho}^2}.
\label{eq:ft}
\ee
The effect of the quantum fluctuations of the atomic positions
on the intersite coupling
here enters through the size of the harmonic oscillator
ground state $a_{\rm ho} = [\hbar/(2m\omega_{\rm ho})]^{1/2}$.

The integral defining $\bar{g}_{\alpha\beta}(\rr)$
is cut at large $k$ 
by the Gaussian factor of momentum
width $1/a_{\rm ho} \ll k_M$. Hence 
we can evaluate the coupling amplitude $\bar{g}_{\alpha\beta}(\RR_i-\RR_j)$
by extending the integral over $\kk$ to the whole space.
In this limit the coupling amplitude is the average,
over the harmonic oscillator ground state probability distributions of $\rr_i$
and $\rr_j$, of the function $g_{\alpha\beta}(\rr_i-\rr_j)$, where 
\cite{pas_delta}
\be
g_{\alpha\beta}(\rr) = -\frac{3\hbar\Gamma}{4k_0^3}\left[
k_0^2\delta_{\alpha\beta}+\partial_{r_\alpha}
\partial_{r_\beta}\right] \frac{e^{ik_0r}}{r}
\ee
is proportional to the component along $\alpha$ of the classical electric field radiated
by a point-like dipole of frequency $\omega_0$ oriented along direction $\beta$
\cite{Morice95}.
A crucial consequence of the average is that, whereas $g_{\alpha \beta}(\rr)$
has the electrostatic $1/r^3$ divergence at the origin, the function
$\bar{g}_{\alpha\beta}(\rr)$ is regular even in $\rr=\mathbf{0}$, with a value
\be
\bar{g}_{\alpha\beta}(\mathbf{0}) = \frac{\hbar\Gamma}{2} \delta_{\alpha\beta}
\left[
\frac{\mbox{Erfi}\,(k_0 a_{\rm ho})-i}{e^{(k_0 a_{\rm ho})^2}}
-\frac{1+2(k_0a_{\rm ho})^2}{2\pi^{1/2} (k_0 a_{\rm ho})^3}
\right]
\label{eq:gbz}
\ee
where $\mbox{Erfi}\,(x)=2\pi^{-1/2}\int_0^x dy\, \exp(y^2)$ is the 
imaginary error function.
For large $r$, the average over the atomic motion
does not suppress the long range nature of the radiated dipolar field
$\propto \exp(i k_0r)/r$.
From (\ref{eq:defu}) one has indeed
$\bar{g}_{\alpha\beta}(\rr)
\simeq  g_{\alpha\beta}(\rr) \exp(-k_0^2 a_{\rm ho}^2)$
remarkably as soon as $r\gg a_{\rm ho}$.

\noindent{\sl Periodic case:} We now take the limit $N\to +\infty$ with
one atom per lattice site, realizing for the light field a periodic potential,
here an arbitrary Bravais lattice.
To diagonalize $H_{\rm eff}^e$
we rely on Bloch theorem: The eigenvectors $|\psi_\qq\rangle$ depend
on the lattice site position $\RR_i$ as
\be
\langle i:e_\alpha||\psi_\qq\rangle = \bar{d}_\alpha e^{i\qq\cdot\RR_i},
\ee
where the Bloch vector $\qq$ 
is chosen in the first Brillouin zone of the lattice.
I.e.\ the ``dipole" carried by atom $i$ differs from the one
$\bar{\mathbf{d}}$ carried by the atom in $\RR=\mathbf{0}$ 
by a global phase factor. Thus the infinite dimension eigenvalue problem
on the excitation spectrum $\varepsilon_{\qq}$,
\be
[H_{\rm eff}^e -E_g P_e] |\psi_\qq\rangle = \varepsilon_{\qq} |\psi_\qq\rangle,
\ee
reduces to the diagonalization of the $3\times 3$ matrix $M$,
$M \bar{\mathbf{d}} = \varepsilon_{\qq} \bar{\mathbf{d}}$, with matrix elements
\be
M_{\alpha\beta} = 
\left(\hbar\omega_A-i\frac{\hbar\Gamma}{2}\right)\delta_{\alpha\beta}
+ \sum_{\RR\in L^*} \bar{g}_{\alpha\beta}(\RR) e^{i\qq\cdot\RR},
\label{eq:spr}
\ee
where the sum runs over the lattice $L$ excluding the origin. 
By adding and subtracting $\bar{g}_{\alpha\beta}(\mathbf{0})$
and using Poisson's summation
formula we convert this sum into a sum over the reciprocal lattice $RL$:
\be
M_{\alpha\beta} = 
\left(\hbar\omega_A-i\frac{\hbar\Gamma}{2}\right)\delta_{\alpha\beta} 
-\bar{g}_{\alpha\beta}(\mathbf{0})
+ \frac{1}{\mathcal{V}_L}\sum_{\KK\in RL} \cheta{\bar{g}}_{\alpha\beta}(\KK-\qq)
\label{eq:pfin}
\ee
where $\mathcal{V}_L$ is the primitive unit cell volume of the lattice.

\noindent {\sl Imaginary part:}
Since each term of the sum over $\KK$ in (\ref{eq:pfin}) is real,
the imaginary part of the excitation spectrum can be calculated explicitly: 
\be
\mbox{Im}\, \varepsilon_\qq = -\frac{\hbar\Gamma}{2} \left(1-e^{-k_0^2 a_{\rm ho}^2}\right).
\label{eq:im}
\ee
As already mentioned, this non-zero value is a direct consequence of the fluctuating
scatterers positions. This is why an expression analogous, but {\sl not equal} to
(\ref{eq:im}) was derived in \cite{Kagan} in the different context of $\gamma$
ray nuclear scattering in a crystal.

The non-zero imaginary part of $\varepsilon_\qq$ 
is indeed due to the decay
of the system out of the subspace where all the atoms
are in their motional ground state.
When an excited atom  $e_\alpha$ in its motional ground state $|\mathbf{0}\rangle_{\rm ho}$ 
spontaneously emits a photon of polarization 
$\epsb$ and momentum $k_0 \mathbf{n}$, $|\mathbf{n}|=1$, 
its probability density to fall in $g$
with an {\sl excited} motional state is \cite{Wineland}
$(3\Gamma/2) |\epsilon_\alpha|^2 
[1-|{}_{\rm ho}\langle \mathbf{0}|\exp(-i k_0 \mathbf{n}\cdot\rrh)|\mathbf{0}\rangle_{\rm ho}|^2]$,
which after sum over $\epsb\perp\mathbf{n}$ and
average over the direction $\mathbf{n}$,
exactly gives the decay
rate $-2\mbox{Im}\, \varepsilon_\qq/\hbar$.
This process conserves the quasi-momentum $\qq$.
If the emitted photon carries away the quasi-momentum $\qq_{\rm ph}$, 
the resulting $g$ atom in the motional excited state 
is coherently delocalized over the whole 
lattice, with a probability amplitude $\propto e^{i\qq_{\rm at}\cdot\RR}$
of being in site $\RR$,
and a quasi-momentum $\qq_{\rm at} = \qq-\qq_{\rm ph}$.
Since $\qq_{\rm ph}$ belongs to a continuum, 
this spontaneous emission process opens up a continuum of final states,
hence the possibility to have for a {\sl fixed} $\qq$
a continuous spectrum for $H$ and a complex $\varepsilon_\qq$.
Experimentally, to obtain long lived elementary excitations,
one may operate in the so-called Lamb-Dicke regime,
$k_0 a_{\rm ho}\ll 1$, where
the loss rate $-2\mbox{Im}\, \varepsilon_\qq/\hbar \simeq
\Gamma (k_0 a_{\rm ho})^2$ is much smaller than  $\Gamma$. 

\begin{figure}[t]
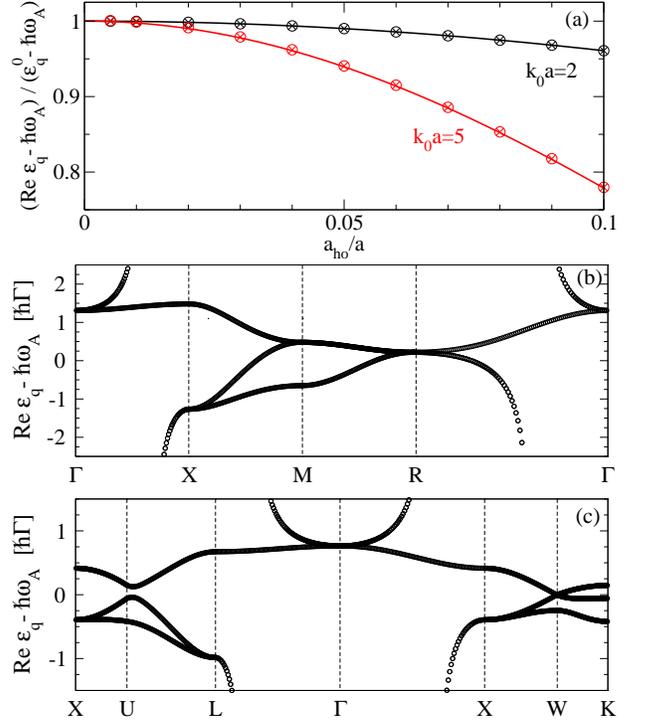

\begin{center}
\includegraphics[width=0.45\textwidth,clip=]{fig1a.eps}
\includegraphics[width=0.45\textwidth,clip=]{fig1bc.eps}
\caption{Spectrum of light in a quantum fluctuating periodic atomic structure. 
(a) For a simple cubic lattice, with a lattice constant $a$ ($\mathcal{V}_L=a^3$),
comparison of the analytical prediction (\ref{eq:jolie}) (solid lines)
with the numerical solution of (\ref{eq:mr}) (symbols),
for two values of $k_0 a$ and of the Bloch vector ($\times$: point $R$
and $\circ$: point $\Gamma$ of the first Brillouin zone).
(b) and (c): Real part of the spectrum  as a function of the Bloch vector
along the standard irreducible path in the first Brillouin zone,
for (b) a simple cubic lattice, and for (c) a face-centered cubic lattice 
with a lattice constant $2a$ ($\mathcal{V}_L=2a^3$);
here $k_0 a=2$ and $k_0 a_{\rm ho}=1/\sqrt{30}$, which leads to
$a_{\rm ho}/a\simeq 0.09$.  
In both cases the band structure is gapless.}
\label{fig:band}
\end{center}
\end{figure}

\noindent{\sl Real part:} 
In (\ref{eq:pfin}) we replace $\bar{g}_{\alpha\beta}(\mathbf{0})$ and 
$\cheta{\bar{g}}_{\alpha\beta}$ by their explicit expressions
(\ref{eq:gbz}) and (\ref{eq:ft}). Then $\mbox{Re}\,\varepsilon_{\qq}-\hbar\omega_A$
is an eigenvalue of the $3\times 3$ real symmetric matrix
\begin{multline}
\delta M_{\alpha\beta} = 
\frac{\hbar\Gamma}{2} \delta_{\alpha\beta}      
\left[
\frac{1+2(k_0a_{\rm ho})^2}{2\pi^{1/2} (k_0 a_{\rm ho})^3} 
-\mbox{Erfi}\,(k_0 a_{\rm ho})\, e^{-k_0^2 a_{\rm ho}^2}
\right]  \\
+
\frac{3\pi\hbar\Gamma}{k_0^3 \mathcal{V}_L}
\sum_{\KK\in RL} \frac{\delta_{\alpha\beta}K'^2-K'_\alpha
K'_\beta}{k_0^2 -K'^2}\,
e^{-K'^2 a_{\rm ho}^2}
\label{eq:mr}
\end{multline}
where $\KK'\equiv \KK-\qq$ \cite{divergences}.
Eq.(\ref{eq:mr}) is useful for a numerical calculation of the spectrum,
see Fig.\ref{fig:band}.
Remarkably, one can even derive analytically the dependence
of the spectrum with $a_{\rm ho}$ from (\ref{eq:spr}):
one multiplies $\bar{g}_{\alpha\beta}(\RR)$ by $e^{k_0^2 a_{\rm ho}^2}$
and one takes the derivative with respect to $a_{\rm ho}^2$. In the Fourier
representation (\ref{eq:ft}), this pulls out a factor $k_0^2-k^2$ which exactly cancels
the denominator. The resulting Fourier integral is now essentially Gaussian
and can be directly calculated:
\begin{multline}
u_{\alpha\beta}(\rr) \equiv
\partial_{a_{\rm ho}^2} \left[e^{k_0^2 a_{\rm ho}^2} \,
\bar{g}_{\alpha \beta} (\rr)\right]
=
\frac{3\hbar\Gamma\, e^{k_0^2 a_{\rm ho}^2}}{8\pi^{1/2}(k_0 a_{\rm ho})^3} \\
\times \left(-\delta_{\alpha\beta} \Delta_{\rr} + \partial_{r_\alpha}
\partial_{r_\beta}\right)
e^{-r^2/(4 a_{\rm ho}^2)}.
\label{eq:defu}
\end{multline}
From (\ref{eq:spr}) one then obtains a sum excluding the origin:
\be
\partial_{a_{\rm ho}^2} \left[e^{k_0^2 a_{\rm ho}^2}\, \delta M_{\alpha\beta} \right]
=\sum_{\RR\in L^*} u_{\alpha\beta}(\RR) \, e^{i\qq\cdot\RR}.
\label{eq:deriv}
\ee
It is apparent that each term of the sum is exponentially small
since, in the deep lattice limit, the harmonic length $a_{\rm ho}$ is much smaller
than the lattice spacing. Keeping these terms is actually beyond accuracy
of our Hamiltonian (see \cite{pas_delta}).
One can thus set the right-hand side of (\ref{eq:deriv}) to zero, which leads to
$\delta M_{\alpha \beta} =  e^{-k_0^2 a_{\rm ho}^2}\, \delta M_{\alpha \beta}^{0}$,
where $\delta M_{\alpha \beta}^{0}$ is independent of $a_{\rm ho}$ and
is simply the limit of $\delta M_{\alpha \beta}$ when $a_{\rm ho}\to 0$.
This shows that the real part of the excitation spectrum 
is a Gaussian function of $a_{\rm ho}$:
\be
\mbox{Re}\, \varepsilon_\qq -\hbar \omega_A = 
e^{-k_0^2 a_{\rm ho}^2}\,
\left(\varepsilon_\qq^0 -\hbar \omega_A \right),
\label{eq:jolie}
\ee
where $\varepsilon_\qq^0$ 
is the limit of the excitation energy for $a_{\rm ho}\to 0$.
Eq.(\ref{eq:jolie}) is the main result of this work, it gives in a very explicit way
the influence of the fluctuations of the atomic positions on the real part of the light
spectrum. It is in excellent agreement with the numerics, see Fig.\ref{fig:band}a.

\noindent{\sl Band structures:}  
We illustrate (\ref{eq:mr}) by numerically calculating the band structure 
in the two lattice geometries of \cite{Lagendijk96,Knoester06}:
In Fig.\ref{fig:band}
we show $\mbox{Re}\,\varepsilon_{\qq}-\hbar\omega_A$ as a function of the Bloch
vector along the standard irreducible path in the first Brillouin
zone, for the simple cubic (Fig.\ref{fig:band}b) and
for the face-centered cubic ({\sl fcc}) (Fig.\ref{fig:band}c) lattices.
The figure reveals the lack of an omnidirectional gap, 
which confirms the prediction of \cite{Knoester06}
against the one of \cite{Lagendijk96} for fixed atomic positions \cite{paper_gap}.
We also explored the {\sl bcc} and several generic less symmetric Bravais lattices without
finding an omnidirectional gap. On the contrary, the non-Bravais 
atomic diamond lattice may support an omnidirectional gap \cite{paper_gap}.
Note that, according to (\ref{eq:jolie}), changing $a_{\rm ho}$ amounts
to a mere rescaling of the vertical axis of Fig.\ref{fig:band}b,c
and cannot open or close an energy gap. The situation may be different
for anisotropic microtraps.

\noindent{\sl Experimental issues:}
The perturbative regime $\Gamma\ll \omega_{\rm ho}$ considered here
was not usual in atomic lattice experiments.
It is now available in experiments using very narrow transitions
for atomic clock purposes:
e.g.\ ${}^{88}$Sr was recently trapped in a deep 3D lattice,
with $a_{\rm ho}\sim 0.05 a\sim 20$nm
\cite{Katori2008}, and it has
a narrow line $5s^2\ {}^{1}S_0\to5s5p\ {}^3P_1$
realizing the needed $J=0\to J'=1$ transition, with $\Gamma\sim 0.05 \omega_{\rm ho}$.
To produce and spectroscopically probe elementary excitations with
$q \neq k_0$, one cannot use
the direct $g\to e$ coupling with resonant light,
but one can use an indirect Raman coupling \cite{Arimondo}.

We also assumed that $g$ and the three sublevels of $e$
experience the same trapping potential.
This is the case in Wigner ion crystals \cite{Wineland95}, 
where $e$ and $g$ experience the same Coulomb shift.
For neutral atoms, the lattice potential
is a lightshift, that may deviate from one of the two
assumptions of (\ref{eq:H0}), 
(i) $e$ experiences a {\sl scalar} lightshift,
and (ii) the lightshifts of $e$ and $g$ are equal.
For a lattice obtained by incoherent superposition of
laser standing waves along $x$, $y$, $z$ linearly polarized
along $y$, $z$ and $x$ respectively,
violation of (i) breaks the harmonic oscillator isotropy:
the sublevel $e_x$ (resp.\ $e_y, e_z$) has an oscillator length
$\eta_e a_{\rm ho}$ 
along $z$ (resp.\ $x,y$) different from the one $a_{\rm ho}$ along the other two directions.
This does not break the three-fold
degeneracy of the motional ground state in $e$ but it reduces 
the overlap between the motional ground state
of $e_\alpha$ and that of $g$. Hence after spontaneous emission, {\sl even}
if one neglects the atom recoil ($k_0 a_{\rm ho}\to 0$), 
the atom in $g$ can populate an excited motional state, giving
a non-zero decay rate to the elementary excitations.
Similarly, violation of condition (ii) leads to an oscillator
length in $g$ equal to $\eta_g a_{\rm ho}$, $\eta_g \neq 1$, which
also increases the decay rate. For $k_0 a_{\rm ho}\to 0$,
combining both violations gives
\be
\mbox{Im}\, \varepsilon_\qq
= -\frac{\hbar \Gamma}{2} \left(1-\frac{8}{[\eta_e\eta_g+(\eta_e \eta_g)^{-1}]
(\eta_g+\eta_g^{-1})^2}\right).
\nonumber
\ee
E.g.\ if one has achieved $\eta_e\simeq 1$ at the expense of having
an optical lattice depth in $g$ twice as small/large as in $e$, one still finds
a small decay rate $\simeq 0.04\Gamma$.

\noindent{\sl Conclusion:} 
Quantum fluctuations of the positions of the scatterers
in a periodic structure very generally give rise to an imaginary part in the spectrum of light
and affect its real part.
For scatterers tightly trapped in a periodic potential, 
we derived an expression for this spectrum.
We showed that, amazingly, its dependence on the amplitude of the 
fluctuations of the positions 
is a Gaussian, not only for the imaginary part 
\cite{Kagan} but also for the real part.
This effect on the real part 
can be large and may be observed in recent atomic lattice clock
experiments.
An intriguing perspective is the extension of
this work to the disordered case and to localized states of light.

We acknowledge discussions with I.\ Carusotto, D.\ Wilkowski, E.\ Arimondo,
D.\ Basko, G. La Rocca, A. Sinatra, and support from IFRAF and ANR Gascor.


\begin{thebibliography}{99}
\bibitem{pas_delta} 
We omitted here contributions in 
$\delta(\rr)$
where $\rr$ is a relative vector between
the positions of two atoms \cite{Morice95}. In a lattice, these contact terms give contributions
$\exp[-(\RR_i-\RR_j)^2/(4 a_{\rm ho}^2)]$ of the order of the tunneling amplitude,
hence totally negligible in the present deep-lattice limit.
\bibitem{neglige}
In the denominator of the $d^2$ term, atomic
motion energy is neglected since it is non-relativistic, $\omega_{\rm ho} \ll \omega_0$.
\bibitem{divergences}
Divergences appear in the
spectrum when $\qq$ is at a distance $k_0$ from a vector $\KK$ 
of the reciprocal lattice. 
This is an artefact of our perturbation theory, which is actually meaningful
only for $|\mbox{Re}\, \varepsilon_\qq - \hbar\omega_A| \ll \hbar \omega_{\rm ho}$.
\end{thebibliography}
\end{document}